\documentclass[aps,floatfix,showpacs,twocolumns,preprintnumbers,amsmath,amssymb]{revtex4}
\usepackage{natbib}
\usepackage{dcolumn}
\usepackage{graphicx}

\newcommand{\be}{\begin{equation}}
\newcommand{\ee}{\end{equation}}
\newcommand{\bea}{\begin{eqnarray}}
\newcommand{\eea}{\end{eqnarray}}

\def\[{\begin{equation}}
\def\]{\end{equation}}
\begin{document}
\title{Contiguous redshift parameterizations of the growth index}
\author{Mustapha Ishak\footnote{Electronic address: mishak@utdallas.edu}}
\author{Jason Dossett\footnote{Electronic address: jnd041000@utdallas.edu}}
\affiliation{
Department of Physics, The University of Texas at Dallas, Richardson, TX 75083, USA}
\date{\today}
\begin{abstract}
The growth rate of matter perturbations can be used to distinguish between different gravity theories and to distinguish between dark energy and modified gravity at cosmological scales as an explanation to the observed cosmic acceleration. We suggest here parameterizations of the growth index as functions of the redshift. The first one is given by $\gamma(a)=\tilde\gamma(a)\,\,\frac{1}{1+(a_{_{ttc}}/a)}+\gamma_{_{early}}\,\,\frac{1}{1+(a/a_{_{ttc}})}$ that interpolates between a low/intermediate redshift parameterization $\tilde\gamma(a)=\gamma_{_{late}}(a)= \gamma_0 + (1-a) \gamma_a$ and a high redshift $\gamma_{_{early}}$ constant value. For example, our interpolated form $\gamma(a)$ can be used when including the CMB to the rest of the data while the form $\gamma_{_{late}}(a)$ can be used otherwise. It is found that the parameterizations proposed achieve a fit that is better than $0.004\%$ for the growth rate in a $\Lambda$CDM model, better than $0.014\%$ for Quintessence-Cold-Dark-Matter (QCDM) models, and better than $0.04\%$ for the flat Dvali-Gabadadze-Porrati (DGP) model (with $\Omega_m^0=0.27$) for the entire redshift range up to $z_{_{CMB}}$. We find that the growth index parameters $(\gamma_0,\gamma_a)$ take distinctive values for dark energy models and modified gravity models, e.g. $(0.5655,-0.02718)$ for the $\Lambda$CDM model and $(0.6418,0.06261)$ for the flat DGP model. This provides a means for future observational data to distinguish between the models.  
\end{abstract} 
\maketitle
\section{introduction}
Cosmic acceleration can be caused by a dark energy component in the universe or a modification to the Einstein field equations of General Relativity at cosmological scales. The growth rate of matter perturbations has been the subject of much recent interest in the literature as a way to distinguish between one possibility or the other, see for example \cite{lue,Aquaviva,gong08b,polarski,linder,Koyama,Koivisto,Daniel,knox,ishak2006,laszlo,Zhang,Hu} for a partial list.  Indeed, distinct gravity theories may have degenerate expansion histories but can be distinguished by their growth rate functions. 
   
As usual, the large scale matter density perturbation $\delta=\delta\rho_m/\rho_m$ satisfies, to linear order, the differential equation
\be
\label{eq:ODE}
\ddot{\delta}+2 H \dot{\delta} -4 \pi G_{eff} \rho_m \delta =0, 
\ee
where $H$ is the Hubble parameter and the effect of the underlying gravity theory is introduced via the expression for $G_{eff}$. The distinct behavior of $\delta$ for different gravity models can be seen in some of the aformentioned references such as for example \cite{ishak2006,linder}.  Equation (\ref{eq:ODE}) can be written in terms of the logarithmic growth rate $f=d \ln{\delta}/d\ln{a}$ as 
\be
\label{eq:ODEf}
f'+f^2+\left(\frac{\dot{H}}{H^2}+2\right)f=\frac{3}{2}\frac{G_{eff}}{G}\Omega_m,
\ee
where primes denote $d/d\ln{a}$.  Throughout this work we will use the numerically integrated solution to this equation normalized at $a = 0\ (z = \infty)$.  Next, the growth function $f$ is usually approximated using the ansatz \cite{Peebles,Fry,Lightman,wang}
\be
f=\Omega_m^\gamma
\ee
where $\gamma$ is the growth index parameter. Reference \cite{Peebles} made an approximation that applies to matter dominated models and proposed $f(z=0)=\Omega_{m0}^{0.6}$ and was followed by a more accurate approximation $f(z=0)=\Omega_{m0}^{4/7}$ in \cite{Fry,Lightman}. Reference \cite{wang} considered dark energy models with slowly varying equation of state, $w$, and found an expression for $\gamma$ as function of $\Omega_m$ and $w$. This has been discussed further in more recent references, see for example \cite{linder07,gong08b}, and also expanded to models with curvature in \cite{Gong09} and \cite{Mortonson}. 

The approaches of expanding the growth index around some asymptotic value or early, matter dominated times with $\Omega_m \approx 1$, or those considering specific redshift ranges to approximate $\gamma$ do not cover other redshift ranges of interest where observational data is available and can constrain the growth parameters or break degeneracies between them and other cosmological parameters. 

Some observational data is already available over the redshift range $z=0-3.8$ \cite{porto,ness,guzzo,colless,tegmark,ross,angela,mcdonald,viel1,viel2} and some recent papers have put some constraints on the values of a constant growth index parameter, see for example \cite{Rapetti2008} where using 
the cosmic microwave background (CMB), type Ia supernovae (SNIa), and X-ray cluster gas-mass fractions ($f_{\rm gas}$), the authors
found $\gamma=0.51^{+0.16}_{-0.15}$ and $\Omega_{\rm m}=0.274^{+0.020}_{-0.018}$ (68.3 per cent confidence limits), for
a flat $\Lambda$CDM background. 
Also, reference \cite{Xia} considered early dark energy (EDE) models and combined data from 
WMAP five-year data release, baryon acoustic oscillations and
type Ia supernovae luminosity distances,  measurements of the linear
growth factors, Gamma-Ray Bursts (GRBs) and Lyman-$\alpha$ forest, 
obtaining $\gamma=0.622\pm0.139\,$ (1$\sigma$) error bar,
which is in agreement with the $\Lambda$CDM and with the values
obtained by other papers \cite{ness,porto}, and as stated there, with 
slightly smaller error bars. Their error bars on $\gamma$ were similar to those
forecasted for future weak lensing and SNIa data by \cite{Hute,linder07}.

In this analysis, we propose parameterizations of the growth that are function of the redshift, covering a wide range of low and intermediate redshifts, and then transitions after some redshift to an almost constant growth index at very high redshifts up to $z_{_{CMB}}$ at the decoupling epoch.  
\section{Interpolated parameterization of the growth index}
First, we recall here the work of references \cite{polarski,gannouji} where the authors proposed a redshift dependent parameterization of the growth index that was intended for the redshift range $0<z<0.5$ \cite{polarski,gannouji} and reads   
\be
\gamma(z)=\gamma_0 +  \gamma'\,z
\label{eq:param1}
\ee
where $\gamma'\equiv \frac{d\gamma}{dz}(z=0)$. This showed already the potential of a variable growth index to distinguish between dark energy models and modified gravity models \cite{polarski,gannouji}. However, current growth data is already available over the redshift range $z=0-3.8$ \cite{porto,ness,guzzo,colless,tegmark,ross,angela,mcdonald,viel1,viel2}, well beyond the $z \ll 1$ approximation. 

In order to be able to consider constraints from this higher redshift data, future growth data that spans over higher redshift ranges, and also very high redshifts up to the CMB scale  (whether to break parameter degeneracies or to put direct constraints), we propose a parameterization that covers such wide ranges and interpolates to an almost constant value of $\gamma$ at very high redshifts up to $z_{_{CMB}}$. Similar to the interpolation proposed in \cite{WMAP5} (see appendix C there) for the equation of state of dark energy, we propose here the following parameterization for the growth index:
\be
\gamma(a)= \tilde\gamma(a)\,h({a}/{a_{_{ttc}}})+\gamma_{early}\,(1-h({a}/{a_{_{ttc}}}))
\ee
where the subscript $ttc$ stands \textit{for transition to a constant} (or almost constant) early growth index, $\gamma_{early}$. The function $h$ is chosen to have the following property  
\bea
h (a/a_{_{ttc}}) \rightarrow  0 &for& a \ll a_{_{ttc}}=1/(1+z_{_{ttc}}) \nonumber\\ 
h (a/a_{_{ttc}}) \rightarrow  1 &for& a \gg a_{_{ttc}}=1/(1+z_{_{ttc}}).
\eea
For simplicity, we adopt the interpolating function \cite{WMAP5} that achieves the behavior above and that is given by 
\be
h(x)=\frac{1}{2}[\tanh(\ln(x)+1]=\frac{x}{x+1},
\ee
and propose the following form for the index parameterization 
\be
\gamma(a)=\tilde\gamma(a)\,\,\frac{1}{1+(a_{_{ttc}}/a)}+\gamma_{_{early}}\,\,\frac{1}{1+(a/a_{_{ttc}})}
\ee 
so that $\gamma(a)$ interpolates between the asymptotic $\gamma_{_{early}}$ value at high redshifts ($z \gg z_{ttc}$) and
\be
\gamma(a)_{_{late}}=\tilde\gamma(a)= \gamma_0 + (1-a) \gamma_a 
\ee
at lower redshifts ($z \ll z_{ttc}$). Similarly, using $a=1/(1+z)$, our parameterization reads
\be
\label{eq:GammaZ}
\gamma(z)=\tilde\gamma(z)\,\,\frac{1}{1+ \frac{{1+z}\,\,\,\,\,\,\,}{1+z_{_{ttc}}}} +\gamma_{early}\,\,\frac{1}{1+ \frac{1+z_{_{ttc}}}{{1+z}\,\,\,\,\,\,\,}}
\ee 
and interpolates between $\gamma_{early}$ (note that this was also noted in the literature as $\gamma_{\infty}$) at high redshift ($z>z_{ttc}$) up to the CMB scale and the following form at lower redshifts, i.e. for $z< z_{ttc}$ 
\be
\label{eq:GammaLateZ}
\gamma(z)_{late}=\tilde\gamma(z)= \gamma_0 + \Big{(} \frac{z}{1+z} \Big{)}\,\, \gamma_a
\ee

It is worth clarifying that the $z_{ttc}$ here is the redshift of transition from a varying growth index parameter $\gamma(z)$ to an almost constant one, i.e. $\gamma_{early}$ (or $\gamma_{\infty}$). This is not necessarily the same $z_{trans}$ that characterizes the transition from a decelerating cosmic expansion to an accelerating one.

We show in the next sections that the proposed parameterizations fit very well the growth function that is numerically integrated from the differential equation (\ref{eq:ODEf}) for a given theory. The fit is better then $0.004\%$ for the $\Lambda CDM$ model for the entire range of redshift from 0 to the $z_{_{CMB}}=1089$ and better than $0.04\%$ for the flat DGP model with $\Omega_m^0=0.27$. We discuss application to these and other models in the next sections. 

\section{Dark Energy Models}
\begin{figure}
\begin{center}
\begin{tabular}{|c|c|}
\hline
{\includegraphics[width=2.8in,height=2.0in,angle=0]{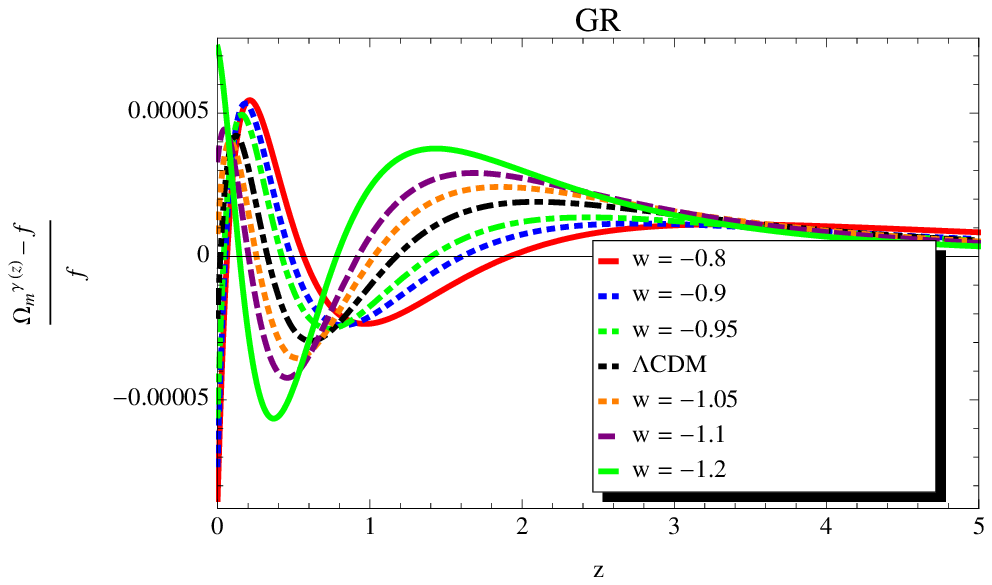}} &
{\includegraphics[width=2.8in,height=2.0in,angle=0]{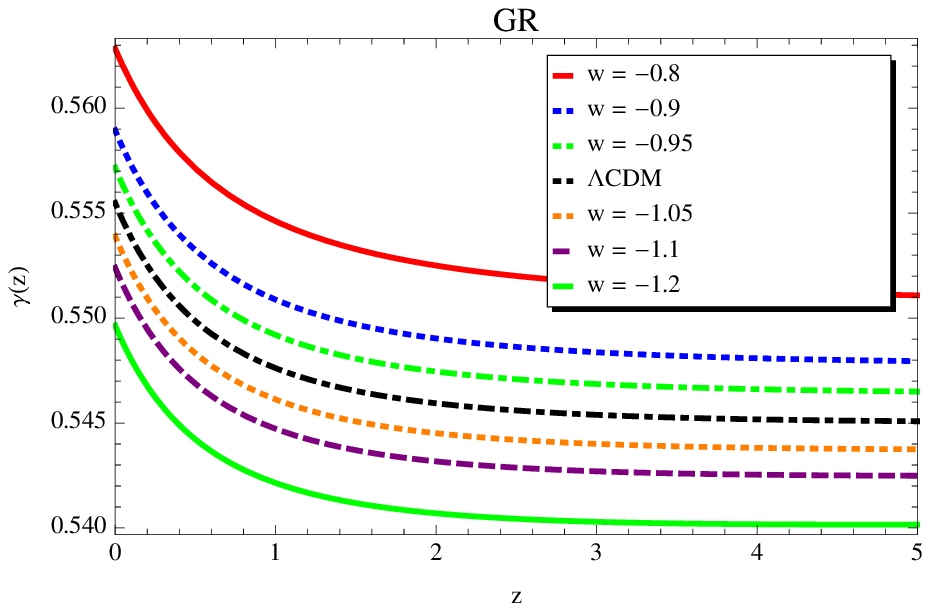}} \\
\hline
{\includegraphics[width=2.8in,height=2.0in,angle=0]{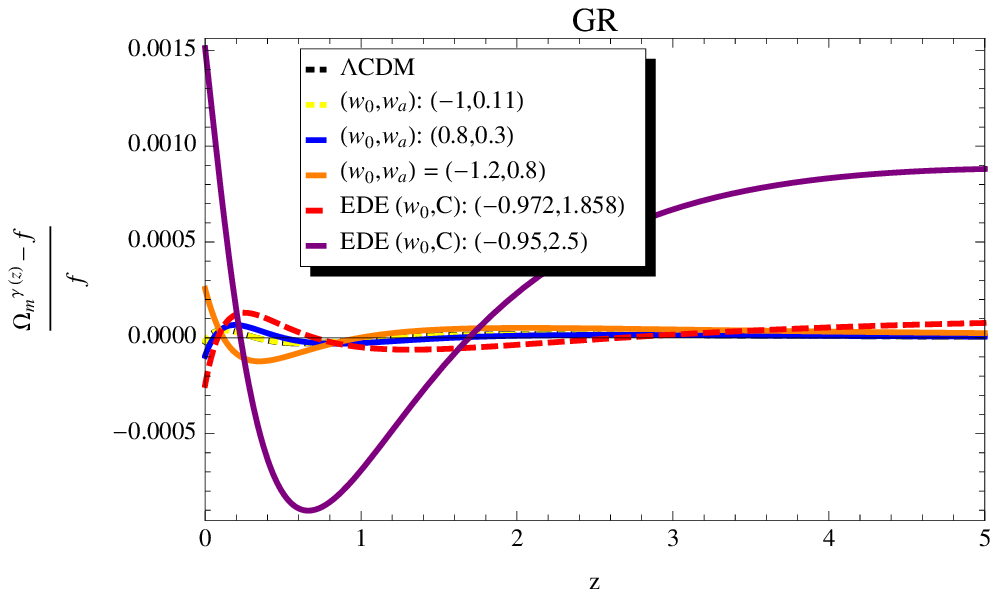}} &
{\includegraphics[width=2.8in,height=2.0in,angle=0]{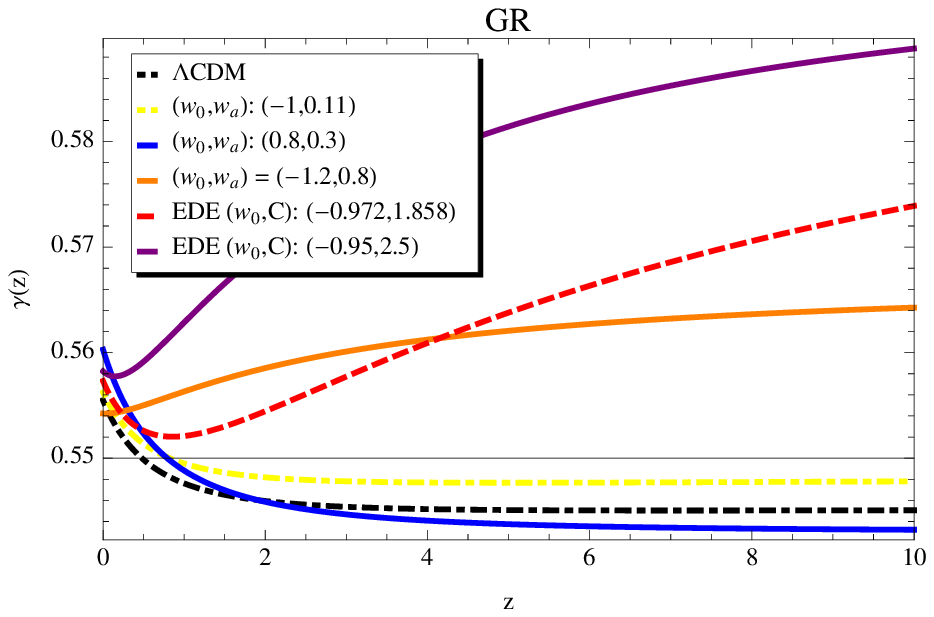}} \\
\hline
\end{tabular}
\caption{\label{fig:GRWfge}
GR - Dark Energy Models. TOP LEFT: We consider the QCDM models with a constant equation of state and plot the relative error $\frac{\Omega_m^{\gamma(z)}-f}{f}$ in order to compare the fit of the proposed parameterization to that of the growth rate, $f$, that is numerically integrated from the growth ODE. For the $\Lambda$CDM, we find the best fit parameters $\gamma_{0}=0.5655$ and $\gamma_a=-0.02710$ when $\gamma_{\infty}^{\Lambda CDM}=6/11$. The fit approximate the growth function $f$ to better than $0.004\%$ while the best fit constant $\gamma_{const}^{\Lambda CDM}=0.5509$ approximates the growth to $0.6\%$. Using our redshift dependent parameterizations of growth index provides an improvement to the fit of the growth of about a factor 150. TOP RIGHT: We plot $\gamma(z)=\tilde\gamma(z)\,\,\frac{1}{1+ \frac{{1+z}\,\,\,\,\,\,\,}{1+z_{_{ttc}}}} +\gamma_\infty\,\,\frac{1}{1+ \frac{1+z_{_{ttc}}}{{1+z}\,\,\,\,\,\,\,}}$ for various values of the constant equation of state $w$ showing very little dispersion of the order of 0.015 at any given redshift. BOTTOM LEFT:We consider the QCDM models with a variable equation of state, as well as some Early Dark Energy models and plot the relative error $\frac{\Omega_m^{\gamma(z)}-f}{f}$ in order to compare the fit of the proposed parameterization to that of the growth rate, $f$, that is numerically integrated from the growth ODE.  We find using our redshift dependent parameterizations of the growth index are able to approximate the growth to within $0.15\%$.  BOTTOM RIGHT:  We plot $\gamma(z)=\tilde\gamma(z)\,\,\frac{1}{1+ \frac{{1+z}\,\,\,\,\,\,\,}{1+z_{_{ttc}}}} +\gamma_\infty\,\,\frac{1}{1+ \frac{1+z_{_{ttc}}}{{1+z}\,\,\,\,\,\,\,}}$ for various dark dnergy models with a varying equation of state $w(a)$ including some early dark energy models.
} 
\end{center}
\end{figure}
For the spatially flat dark energy models with constant equation of state $w$, the Friedmann equations give
\begin{equation}
\label{wcdmhdot}
\frac{\dot H}{H^2}=-\frac{3}{2}[1+w(1-\Omega_m)].
\end{equation}
and energy conservation equation read  
\begin{equation}
\label{wcdmwmder}
\Omega_m'=3w\Omega_m(1-\Omega_m).
\end{equation}
Now, substituting Eqs. (\ref{wcdmhdot}) and (\ref{wcdmwmder}) into Eq. (\ref{eq:ODEf}), one gets
\begin{equation}
\label{wcdmfeq}
3w\Omega_m(1-\Omega_m)\frac{df}{d\Omega_m}+f^2+\left[\frac{1}{2}-\frac{3}{2}w(1-\Omega_m)\right]f=\frac{3}{2}\Omega_m.
\end{equation}
Next, using $f=\Omega_m^\gamma$ into Eq. (\ref{wcdmfeq}) yields
\begin{equation}
\label{wcdmfeq1}
3w\Omega_m(1-\Omega_m)\ln\Omega_m\frac{d\gamma}{d\Omega_m}-3w\Omega_m(\gamma-1/2)+\Omega_m^\gamma-\frac{3}{2}\Omega_m^{1-\gamma}
+3w\gamma-\frac{3}{2}w+\frac{1}{2}=0.
\end{equation}
An expansion of Eq. (\ref{wcdmfeq1}) around $\Omega_m=1$ (early times), to the first order
of $(1-\Omega_m)$, one gets \cite{wang,note}
\begin{equation}
\label{wcdmr}
\gamma=\frac{3(1-w)}{5-6w}+\frac{3}{125}\frac{(1-w)(1-3w/2)}{(1-6w/5)^2(1-12w/5)}(1-\Omega_m).
\end{equation}
Eq. (\ref{wcdmr}) gives the asymptotic expression for very high redshifts  
\begin{equation}
\label{wcdm0}
\gamma_\infty=\frac{3(1-w)}{5-6w}
\end{equation}
which reduces to the well-known $\gamma_\infty^{^{\Lambda CDM}}=\frac{6}{11}$ for the $\Lambda$CDM model so that our parameterization in this case takes the form 
\be
\label{eq:GammaLCDMZ2}
\gamma(z)=\tilde\gamma(z)\,\,\frac{1}{1+ \frac{{1+z}\,\,\,\,\,\,\,}{1+z_{_{ttc}}}} +\gamma_\infty^{^{\Lambda CDM}}\,\,\frac{1}{1+ \frac{1+z_{_{ttc}}}{{1+z}\,\,\,\,\,\,\,}}
\ee 
with $\tilde\gamma(z)$ given by our low/intermediate redshift parameterization (\ref{eq:GammaLateZ}).

We show in Figure \ref{fig:GRWfge} how well various parameterizations of the growth fit the growth rate function $f$ that is integrated numerically from the differential equation (\ref{wcdmfeq}) by plotting the relative error $\frac{\Omega_m^{\gamma(z)}-f}{f}$. For $\Lambda$CDM, we plot the figures up to a redshift of 5 but we performed best fits of the parameters up to the $z_{_{CMB}}\approx1089$.  We find the best fit parameters $\gamma_{0}=0.5655$ and $\gamma_a=-0.02718$ when $\gamma_{\infty}^{\Lambda CDM}=6/11$. The fit approximate the growth function $f$ to better than $0.004\%$ while the best fit constant $\gamma_{const}^{\Lambda CDM}=0.5509$ approximates the growth rate function to about $0.6\%$. Thus using our redshift dependent parameterizations for the growth index gives an improvement to the fit of the growth rate function of about a factor 150. We also plot $\gamma(z)$ for various  dark energy equations of state, including some early dark energy models (see for example \cite{Xia,lindermock,lindermock2,Hollenstein2009} for a discussion of the latter and our appendix for a parameterization). In Table \ref{table:param} we list the best fit parameter values of $\gamma_0$ and $\gamma_a$ for the various models used. We find that these best fit values for $\gamma_0$ and $\gamma_a$ do not change
for a wide range of $z_{ttc}$ from for example 0.5 to several.  

\section{DGP model}
For the spatially flat DGP \cite{dgp} model, the effective gravitational constant is given by 
\begin{equation}
\label{dgpgeff}
\frac{G_{eff}}{G}=\frac{2(1+2\Omega_m^2)}{3(1+\Omega_m^2)}
\end{equation}
and Friedmann equations yield
\begin{equation}
\label{dgphdot}
\frac{\dot H}{H^2}=-\frac{3\Omega_m}{1+\Omega_m}.
\end{equation}
The conservation of energy gives 
\begin{equation}
\label{dgpwmder}
\Omega_m'=-\frac{3\Omega_m(1-\Omega_m)}{1+\Omega_m}
\end{equation}
where the matter energy density is given by
\begin{equation}
\label{dgpwm}
\Omega_m=\frac{\Omega_m^0(1+z)^3}{[(1-\Omega_m^0)/2+
\sqrt{\Omega_m^0 (1+z)^3+(1-\Omega_m^0)^2/4}\,]^2}.
\end{equation}
Now, using equations (\ref{dgpgeff}), (\ref{dgphdot}) and
(\ref{dgpwmder}) into Eq. (\ref{eq:ODEf}), one gets
\begin{equation}
\label{dgpfeq}
-\frac{3\Omega_m(1-\Omega_m)}{1+\Omega_m}\frac{df}{d\Omega_m}+f^2+\frac{2-\Omega_m}{1+\Omega_m}\,f =\frac{\Omega_m(1+2\Omega_m^2)}{1+\Omega_m^2}.
\end{equation}
Next, using $f=\Omega_m^\gamma$ into Eq. (\ref{dgpfeq}), one gets
\begin{equation}
\label{dgpfeq1}
-\frac{3\Omega_m(1-\Omega_m)\ln\Omega_m}{1+\Omega_m}\frac{d\gamma}{d\Omega_m}-\frac{3(1-\Omega_m)\gamma}{1+\Omega_m}
+\Omega_m^\gamma+\frac{2-\Omega_m}{1+\Omega_m}-\frac{\Omega_m^{1-\gamma}(1+2\Omega_m^2)}{1+\Omega_m^2}=0.
\end{equation}
Again, expanding Eq. (\ref{dgpfeq1}) around $\Omega_m=1$, to the first order
in $(1-\Omega_m)$, one gets \cite{gong08b}
\begin{equation}
\label{dgpr}
\gamma=\frac{11}{16}+\frac{7}{5632}(1-\Omega_m).
\end{equation}
So the asymptotic value for the DGP model is $\gamma_\infty^{DGP}=\frac{11}{16}$.
\begin{figure}
\begin{center}
\begin{tabular}{|c|c|}
\hline
{\includegraphics[width=2.8in,height=2.0in,angle=0]{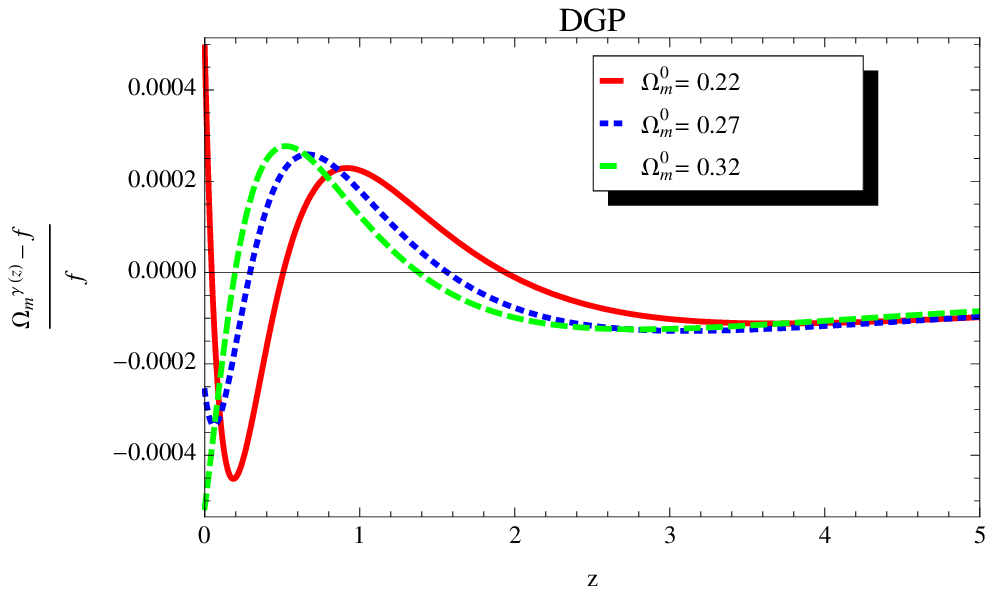}} &
{\includegraphics[width=2.8in,height=2.0in,angle=0]{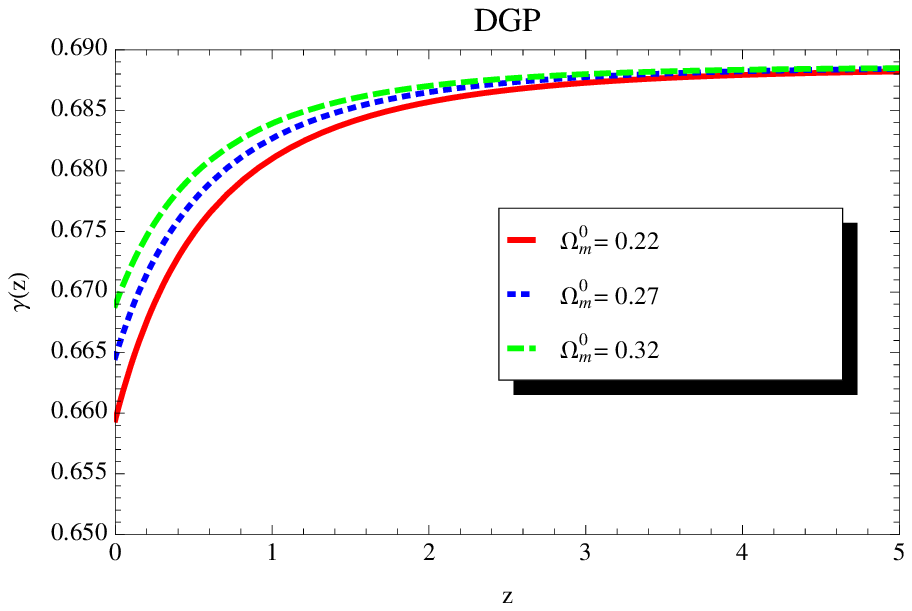}} \\
\hline
\end{tabular}
\caption{\label{fig:DGPWfge}
DGP models. LEFT: We consider the DGP model and plot the relative error $\frac{\Omega_m^{\gamma(z)}-f}{f}$ in order to compare the fit of the proposed parameterization to that of the growth rate $f_{num}$ that is numerically integrated from the growth ODE. We find the best fit parameters $\gamma_{0}=0.6418$ and $\gamma_a=0.06261$ for $\Omega_m^0 = 0.27$ when $\gamma_{\infty}^{DGP}=11/16$. The fit approximates the growth function $f$ to better than $0.04\%$ while the best fit constant $\gamma_{const}^{^{DGP}} = 0.6795$ approximates the growth to $1.95\%$. So using our redshift dependent parameterization of the growth index provides an improvement to the fit of about a factor $50$ for the DGP model. RIGHT: We plot $\gamma(z)=\tilde\gamma(z)\,\,\frac{1}{1+ \frac{{1+z}\,\,\,\,\,\,\,}{1+z_{_{ttc}}}} +\gamma_\infty^{^{DGP}}\,\,\frac{1}{1+ \frac{1+z_{_{ttc}}}{{1+z}\,\,\,\,\,\,\,}}$ for various values of $\Omega_m^0$ showing very little dispersion of the order 0.01 or less at any redshift.  
}
\end{center}
\end{figure}
Using the proposed parameterizations (\ref{eq:GammaZ}) and (\ref{eq:GammaLateZ}) for the DGP model, we show in Figure \ref{fig:DGPWfge} how well the parameterizations fit the growth rate function $f$ that is integrated numerically from the differential equation (\ref{dgpfeq}). For that, we plot the relative error $\frac{\Omega_m^{\gamma(z)}-f}{f}$. We performed the fit for a redshift up to the $z_{_{CMB}}\approx1089$. We find the best fit parameters $\gamma_{0}=0.6418$ and $\gamma_a=0.06261$ for $\Omega_m^0 = 0.27$ when $\gamma_{\infty}^{DGP}=11/16$. The fit approximates the growth function $f$ to better than $0.04\%$ while the best fit constant $\gamma_{const}^{DGP} = 0.6795$ approximates the growth to $1.95\%$. Thus, using our redshift dependent parameterizations of the growth index provides an improvement to the fit of the growth rate function of about a factor $50$ for the DGP model. 
We also plot $\gamma(z)$ using our parameterization for various values of $\Omega_m$ and find that the difference of the order 0.01 or less at any redshift.  In Table \ref{table:param} we list the best fit parameter values of $\gamma_0$ and $\gamma_a$ for the various models used. Again, we find that these best fit values for $\gamma_0$ and $\gamma_a$ do not change
for a wide range of $z_{ttc}$ from for example 0.5 to several.  
\begin{center}
\begin{table}
\begin{tabular}{|c|c|c|}\hline
\multicolumn{3}{|c|}{\bfseries Parameters for various QCDM models.}\\ \hline
$\mathbf{(w_0,w_a)}$&$\mathbf{\gamma_0}$&$\mathbf{\gamma_a}$\\ \hline
$(-0.8,0)$&$0.5690$&$-0.02131$\\ \hline
$(-0.9,0)$&$0.5683$&$-0.022525$\\ \hline
$(-0.95,0)$&$0.5676$&$-0.02699$\\ \hline
$(-1,0)$&$0.5655$&$-0.02718$\\ \hline
$(-1.05,0)$&$0.5635$&$-0.02735$\\ \hline
$(-1.1,0)$&$0.5617$&$-0.02749$\\ \hline
$(-1.2,0)$&$0.5583$&$-0.02771$\\ \hline
$(-1,0.11)$&$0.5641$&$-0.02464$\\ \hline
$(-0.8,-0.3)$&$0.5720$&$-0.03074$\\ \hline
$(-1.2,0.8)$&$0.5409$&$-0.01417$\\ \hline \hline
\multicolumn{3}{|c|}{\bfseries Parameters for some EDE models.}\\ \hline
$\mathbf{(w_0,C)}$&$\mathbf{\gamma_0}$&$\mathbf{\gamma_a}$\\ \hline
$(-0.972,1.858)$&$0.5498$&$-0.02915$\\ \hline
$(-0.95,2.5)$&$0.5165$&$-0.05578$\\ \hline \hline
\multicolumn{3}{|c|}{\bfseries Parameters for various DGP models.}\\ \hline 
$\mathbf{\Omega_m^0}$&$\mathbf{\gamma_0}$&$\mathbf{\gamma_a}$\\ \hline
$0.22$&$0.6314$&$0.07324$\\ \hline
$0.27$&$0.6418$&$0.06261$\\ \hline
$0.32$&$0.6504$&$0.05279$\\ \hline
\end{tabular}
\caption{\label{table:param}
We list the parameter values for in our interpolation parameterization for various QCDM, EDE, and DGP models.  These values were found by fitting our parameterization to the numerically integrated solution of ODE for the growth function, $f$ (e.g. we use for $\gamma(z)$, Eqs.(18) with(9) for dark energy models, and Eqs. (25) with (9) for DGP models). We see that the QCDM and EDE models have a negative values for the parameter $\gamma_a$, while the DGP models have a positive value for $\gamma_a$, thus providing parameter that observational data can constrain to distinguish between the two gravity theories, additionally $\gamma_0$ takes on distinct values for each theory.
}
\end{table}
\end{center}
\section{Conclusion}
Data on growth rate of large scale structure covers already a wide range of redshift
\cite{porto,ness,guzzo,colless,tegmark,ross,angela,mcdonald,viel1,viel2} and is likely to cover even a wider range for incoming and future data. We proposed parameterizations of the growth index that cover such wide redshift ranges and also interpolates to the highest redshifts including the CMB scale. The parameterizations are found to fit the growth function to better than $0.004\%$ over the entire redshift range for the $\Lambda$CDM model,  to better than $0.014\%$ for various QCDM models, and to better than $0.04\%$ for the DGP model (with $\Omega_m^0=0.27$). Such parameterizations should be useful for ongoing and future high precision missions. We find that the best fit values for the growth index parameters take distinctive values for dark energy models versus modified gravity models: $(\gamma_0,\gamma_a)=(0.5655,-0.02718)$ for the $\Lambda$CDM model and $(\gamma_0,\gamma_a)=(0.6418,0.06261)$ for the flat DGP model.  Most notable of the above values is the fact that $\gamma_a$ is of a different sign for the two model. This distinction hold even when looking at more complex dark energy models.  This provides a way for observational data to distinguish between dark energy models and modified gravity models as cause of cosmic acceleration.  

\acknowledgements 
M.I. acknowledges partial support from the Texas Space Grant Consortium, the Hoblitzelle Foundation, and a Clark Award. 
\appendix
\begin{center}
    {\bf APPENDIX}
\end{center}

For early dark energy we use a Mocker model first introduced by \cite{lindermock}. The dark energy equation of state for these models is given by:
\begin{equation}
w(a) = -1+\left[1-\frac{w_0}{1+w_0}a^C\right]^{-1}.
\end{equation}
In these models the dark energy component behaves like nonreletavistic matter at high redshifts, having an equation of state $w = 0$, but assymptotes to a cosmological constant with $w = -1$.  We use parameter values for $w_0$ and $C$ given by \cite{Xia} which are said to fit CMB and SN Ia constraints very well.  See the aformentioned references as well as \cite{lindermock2} for a more in depth description of these models.


\begin{thebibliography}{}
\bibitem{note} We find a small discrepancy in the denominator
of the second term of equation (\ref{wcdmr}) compared to the widely
cited expression for the growth index, i.e. equation (B11) from pioneering
work by \cite{wang}.
\end{thebibliography}
\end{document}